\documentclass[a4paper]{jpconf}
\usepackage{graphicx}
\usepackage{bm}
\usepackage{siunitx}
\begin{document}
\title{Evaluating the charged background rejection requirement 
in an experiment to measure $\rm BR(K_L \to \pi^0 \nu \bar{\nu})$ at the CERN SPS}

\author{A Bradley}

\address{Department of Physics and Astronomy, George Mason University, 4400 University Drive, Fairfax, Virginia 22030, USA}

\ead{abradl12@gmu.edu}

\begin{abstract}
Measuring the rate at which the long-lived, neutral kaon decays into a neutral pion, neutrino, and anti-neutrino allows physicists an opportunity to test precise predictions made by the Standard Model. Differences between theoretical predictions and experimental measurements may point to new physics. Not only does the Standard Model predict a very low probability at approximately 3 $K_L \to \pi^0 \nu \bar{\nu}$ decays in 100 billion $K_L$ decays, but many of the common decays leave false signals in the detector that look the same as the true signal. Charged decays have been studied to determine the required detection efficiency necessary to eliminate them. The conclusion of these studies is that a reduction by a factor of $1/(\num{3e9})$ will be required to achieve the 10:1 signal to charged background ratio necessary for the experiment.
\end{abstract}

\section{Introduction to the ``KLEVER'' feasibility study}

A small collaboration within the NA62 experiment is conducting a feasibility study for measuring the very rare decay $K_L \to \pi^0 \nu \bar{\nu}$. With a BR($3.4\pm0.6)\times10^{-11}$~\cite{Buras} predicted by the Standard Model (SM), no experiment has yet been capable of observing the decay. The goal of the ``KLEVER'' experiment is to observe tens of these decays while keeping the signal to background ratio at 1:1. The collaboration has published their initial findings in an NA62 technical note~\cite{KLEVER}, including the findings of this study.

To accomplish this goal, long-lived, neutral kaons $K_L$ will be created by colliding 400 GeV/c protons accelerated by CERN's SPS with a beryllium target. The most common charged decays from the $K_L$, the signal decay, and their applicable BR's are shown in Table~\ref{table}. The signal is two photons from the $\pi^0$ detected by the liquid krypton electromagnetic calorimeter (LKr) and no other particles detected elsewhere. Any difference between the experimental measurement and the SM branching ratio may point to new physics.\cite{Smith}

\section{Background contribution from charged decays}

A 10:1 signal to charged background ratio is necessary for the experiment. The rate of charged background and the reduction required to achieve the signal to background ratio were studied. The overall 1:1 signal to background ratio includes the neutral decays, with contributions primarily from $K_L \to \pi^0 \pi^0$, but is not discussed here. Additional details can be found in~\cite{KLEVER}.

\subsection{The Detector}

In addition to the LKr ($z = 242$ m from the target), there are 26 large angle vetoes (LAV) upstream of the LKr, and a small angle calorimeter (SAC) and inner ring calorimeter (IRC) behind the LKr. Their purpose is to detect photons and charged particles missed by the LKr.

\subsection{Simulation and Analysis}

Charged decays of the kaon were simulated with \textsc{Geant4}~\cite{Geant4_1}\cite{Geant4_2}. A list of selection criteria was sequentially applied to each event in the resulting \textsc{Root}~\cite{ROOT} file. Events were eliminated if any of the following were true: a charged particle or photon was detected by the LAVs, IRC, or SAC; fewer than two energy clusters were detected by the LKr (clusters closer than 6 cm were merged); the reconstructed $\pi^0$ decayed outside of the fiducial region ($105 < z < 155$ m); an LKr cluster was closer than 35 cm to the beamline center; or the transverse momentum of the reconstructed $\pi^0$ was less than 0.12 GeV/c. Designed primarily to eliminate background from the most problematic neutral decays, these criteria also significantly reduce charged background.

\subsection{Results and Conclusion}

Some charged decays passed all of the selection criteria and were accepted as signal events because they were indistinguishable from the signal (e.g. a pion and electron reconstructed as two photons from a $\pi^0$) [see Table~\ref{table}]. 

\begin{table}[h]
\caption{\label{table}Analysis results for charged and signal decay modes. $\epsilon$ is the probability a generated event was accepted as a signal. Charged BRs are from \cite{PDG} and the signal BR is from \cite{Buras}.}
\begin{center}
\lineup
\begin{tabular}{*{6}{l}}
\br                              
Decay  
	& Generated 
	& Accepted 
	& $\epsilon$ 
	& BR
	& $\epsilon\times{\rm BR}$ \cr 
Mode
	& $(\times10^6)$ 
	& as Signal
	& 
	& 
	& 
	\cr 
\mr
$\pi^{\pm} e^{\mp} \nu_e$ 
	& \0\00.24
	& \0\018
	& $7.4\times10^{-5}$
	& $(40.55 \pm 0.11)$\%   
	& $3.0\times10^{-5}$
	\cr
$\pi^{\pm} \mu^{\mp} \nu_{\mu}$ 
	& \044.8
	& \0\023
	& $5.1\times10^{-7}$
	& $(27.04 \pm 0.07)$\%
	& $1.4\times10^{-7}$
	\cr 
$\pi^{+} \pi^{-} \pi^{0}$ 
	& 153.0
	& \0\011
	& $7.2\times10^{-8}$
	& $(12.54 \pm 0.05)$\%
	& $9.0\times10^{-9}$
	\cr 
$\pi^{+} \pi^{-}$ 
	& \0\00.73
	& \0\016
	& $2.2\times10^{-5}$
	& \0$(1.97 \pm 0.01) \times 10^{-3}$
	& $4.3\times10^{-8}$
	\cr 
$\pi^{0} \nu \bar{\nu}$ 
	& \0\02.8
	& 6678
	& $2.4\times10^{-3}$
	& \0$(3.4\0 \pm 0.6) \times 10^{-11}$
	& $8.1\times10^{-14}$
	\cr
\br
\end{tabular}
\end{center}
\end{table}

The last column of Table~\ref{table} shows that a charged particle detection (CPD) system will need to reduce the charged background from $\num{3e-5}$ to $\num{8e-15}$ to achieve the 10:1 signal to charged background ratio---a reduction by a factor of $1/(\num{3e9})$.

\ack
This work was supported by the U.S. National Science Foundation under Grant No. 1506088 and funding under a grant from the Italian Ministry of Education, Universities, and Research (MIUR) as the “KLEVER” project, a Research Project of Significant National Interest (Progetto di ricerca di Rilevante Interesse Nazionale, PRIN), call 2010-2011.

\end{document}